# Toward Semantic Representation of Middleware Services

Alaa Abd Elhamid Radwan , Mohammad Tabrez Quasim
*Collage of Computing & Information Technology,*
*University of Bisha, Saudi Arabia*
{*ardwan,mtabrez}@ub.edu.sa*

ABSTRACT: Middleware is middle tier software that supports communications between two or more different applications, and between applications and shared services Managing the complexity and heterogeneity of distributed infrastructures is the important role of middleware so that it can provide the simple programing environment for the developer of distributed application Middleware supports communications, information exchange, objects management, sending messages, in addition to provides many functions to build distributed systems. Many classification and definition has been provided for middleware, there is a great need to give a global look to middleware and their factors. Ontology for Middleware services is proposed in this paper, whose intention is to make a global look to middleware and enrich middleware services description. Using this ontology, we can discover and classify incoming services to their appropriate types according to their specification and characteristics. The proposed ontology is helpful for users to find their suitable services according to their own preferences. Besides this is an example, a parser has been used to classify many middleware service files.

## 1 INTRODUCTION

Middleware is a very vague term and there are lots of definitions which try to define the concept. A middleware can be precisely defined as  "Software layer that lies between the operating system and applications on each side of a distributed computing system in a network." (Sacha ,2019) the supported API and protocols by the middleware services are responsible to defined it
Transparency in location is also very vital feature provided by middleware which simply means that the components can travel between the computers without doing any changes in other components (Jean, 2005; Ajay,2008).
A proposed classification of middleware can appear as depicted in (figure 1), These categories are not always clearly isolated from each other, the various types of middleware overlap in some services provided.
   The great benefit of Middleware is to provide API services (through many distributed software) that work between the application and the operating system and network to enable the application to achieve the following goals (Amo,2006;Max, 2006) in table-1

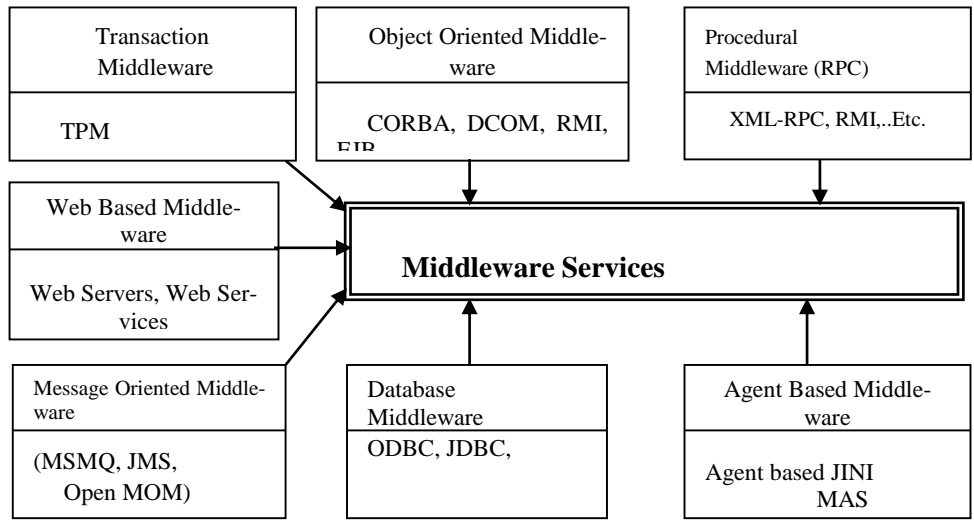

Fig. 1: Middleware Classification and services

Table : 1

| S.No | Goals |
|---|---|
| 1 | Locate across the network transparently, and provide interaction with another application or service. |
| 2 | To become independent from network services. |
| 3 | To become reliable and available |
| 4 | To provide scalability without losing functionality |

## 2. MIDDLEWARE TYPES

Many classifications and definitions have been provided for middleware. There is an urgent need to shed light on middleware services and their factors. Some classification of middleware include (Chris, 2004; Cambel, 1999; Myerson, 2002)
1- Object oriented middleware (distributed object middleware).
2- Remote procedure call.
3- Message oriented middleware.
4- Database middleware
5- Transaction Middleware
6- Agent Based Middleware
7- Web Based Middleware
However this is not the complete list, rather we can also find some other software included in the category , Now we will begin to briefly discuss each type of previous middleware services

## 2.1 OBJECT ORIENTED MIDDLEWARE (OOM)

Object oriented middleware is based on the simple concept of invoking an operation in an object that available in another system, unlike of client and server concept t, in this concept there are client and object(Quasim,2013a;Quasim,2013b). It provides a mechanism to allow methods to be invoked on remote objects, and provide services to support the naming and location of objects in a system-wide manner (usually called Object Directory, Yellow Pages, etc) in addition to inheritance, object references and exceptions. Middleware systems in this category include the OMG's CORBA, Microsoft COM, Java RMI and Enterprise Java Beans (EJB) (ERL,2003; OMG,1995; Siegel,2001)

Table 2: OOM characteristics

| Property | Value |
| --- | --- |
| **Request Reference** | Distributed Object |
| **Connection Point** | Client/Server Stubs |
| **Connection Mode** | Synchronous (mainly) |
|  | Asynchronous (limited) |
| **Scalability** | Limited |
| **Message / Transaction** | Supported |
| **Heterogeneity** | Language – Independent * |

\* CORBA and DCOM support language independent (object written using different languages) where Java/RMI addresses the Virtual Java system heterogeneity

OOM Advantage:
1- Supports both synchronous and asynchronous communication.
2- Marshalling and unmarshalling are generated automatically.
3- Most of the OOM products support the messaging and transactions facilities

OOM Disadvantage:
1- The main disadvantage of OOM is its lack of scalability.
2- Lack of interoperability between different OOM products (CORBA, DCOM … etc).

## 2.2 PROCEDURAL MIDDLEWARE (RPC)

Sun Microsystems developed Remote Procedure Calls (RPCs) in the early 1980's.. RPCs are available on various operating systems, including most UNIX and M.S Windows systems. RPC contains a client program to call procedures located on a remote server program. Stubs are developed for both the client and the server to call up synchronously when the client makes a call to the server.

A web version of RPC is XML-RPC. XML-RPC is a simple, portable way to make remote procedure calls over HTTP. It can be used with many programming languages such as Perl, Java, C, C++, PHP.. etc. Implementations are available for UNIX, Windows and the Macintosh. XML-RPC combining RPC architecture with XML and HTTP technology

RPC Advantages:

1. PCs are naturally supported, because RPC has bindings for multiple operating systems and programming languages.

2. Marshalling and un-marshalling are automatically generated.

RPC Disadvantages:
1.RPCs don't support synchronous communication.
2- Limited scalability because there are no direct support for replication and load balancing.
3- Fault tolerance is worse than by other middleware types.

Table 3: Procedural Middleware characteristics:

| Property | Value |
| --- | --- |
| **Request Reference** | Remote Procedure |
| **Connection Point** | Client/Server Stubs |
| **Connection Mode** | Synchronous |
| **Scalability** | Limited |
| **Client state** | Blocked (Mainly) |
| **Heterogeneity** | Language – Independent |

*2.3 MESSAGE ORIENTED MIDDLEWARE (MOM)*

MOM is another important middleware that establish a communication between applications through messages. MOM allows both type of synchronous and asynchronous communication models leading to two different types of MOM: message queuing and message passing, a third model is also existing known as a "publish and subscribe" model .
- Message passing mechanism support the direct communication between applications..
- Message queuing however supports the indirect communication model and always implies a connectionless way of interacting.
- In pub/sub mode clients have the ability to subscribe to the interested subjects. After subscribing, the client will receive any message corresponding to a subscribed topic.

Table 4: MOM characteristics

| Property | Value |
| --- | --- |
| **Network communication** | Messages |
| **Connection Point** | Client/Server |
| **Connection Mode** | Synchronous<br>Asynchronous |
| **Scalability** | Limited |
| **Heterogeneity** | Limited * |

* Marshalling is not automatically generated ,and programmers must implement it.

MOM Advantages and Disadvantages

| Advantages | 1.MOM supports group communication, which is atomic . |
| --- | --- |
| | 2. Persistent queues improve MOM products reliability . |
| | **3-**Most MOM products support queues for transactional messages |

|  | 4- MOM enhances flexibility. Participants need not to know about the existence of each other. |
|---|---|
| Disadvantages: | 1- MOM has limited scalability and heterogeneity support. |
|  | 2- Bad portability because MOM products are not compatible |

*2.4 TRANSACTION MIDDLEWARE (TM)*

In order support the concept of distributed transaction across different host the Transactional middleware (TM) or transaction processing monitors (TPM) were designed

Although it is compulsory for A transaction that it shall enable ACID properties (Atomic, Consistent, Isolated, and Durable). It usually uses the two-phase commit (2PC) protocol to implement these type of transactions and properties(Speer, 1994; Ozen, 1991)

Table 5: TPM Characteristics:

| Property | Value |
|---|---|
| **Request Reference** | Distributed transactions |
| **Connection Point** | Client/Server Component |
| **Connection Mode** | Synchronous / Asynch. |
| **Scalability** | High |
| **Client state** | Blocked (Mainly) |
| **Heterogeneity** | Medium * |

" *TM supports soft- and hardware heterogeneity, but doesn't support data heterogeneity very well, because it can't express complex data structures and therefore can't marshals these structures."

TPM Advantages:

| |
|---|
| **1-** Components are maintained in consistent states because of the transaction's ACID properties. |
| **2**- TM is very trustworthy |
| **3**- Supports both synchronous and asynchronous communication between hosts. |
| 4- TP monitors can dispatch, schedule and prioritize multiple application requests concurrently. |

TPM Disadvantages:

| |
|---|
| 1.Transactions have a significant overhead to manage. |
| 2- The guarantees they provide according to ACID properties are often unnecessary or undesirable. |
| 3- Most transactional middleware systems do not provide any automated marshalling or unmarshalling. |
| 4- TM runs on fewer platforms as compared to other types of middleware. |

| | |
|---|---|
| 5- Long-lived activation transactions could prevent continues of other clients. | |

## 2.5 DATABASE MIDDLEWARE

Database middleware is responsible to establish communication among different applications and Local/remote database. However the database middleware cannot allow two way communication as well as transfer calls/objects between client and server(Linthicum,2019)

It is often chosen to complement other middleware. types of Database middleware may be varying according to data source (flat file, relational DB, object database...), or connection type such as native middleware, call level interfaces (CLIs) and database gateways .

Native middleware is a middleware created for a specific database, it provides the best performance and access to native database features. Call Level Interface or CLIs provide a single interface to several databases. CLIs, such as OLE DB, ODBC and JDBC, provide a single interface to several databases.

They are capable to translate the common interface call
into number of database dialects, and also translate the response back in an understandable form to the requesting application ( Microsoft Corporation.1992)

Database gateways is capable to integrate different databases for access from a single application interface.

Table 6: Database Middleware Characteristics:

| Property | Value |
|---|---|
| **Request Reference** | SQL Query |
| **Connection Point** | Client/Server |
| **Connection Mode** | Synchronous |
| **Scalability** | High |
| **Client state** | Blocked (Mainly) |
| **Heterogeneity** | High |

Database Middleware Advantages:

| |
|---|
| 1. Provides access to nearly all types of databases. |
| 2- Facilitates communication among applications and local or remote databases. |
| 3- Provides support for transactional operations. |
| 4- Normally chosen to incorporate other types of middleware. |
| 5- Works as a translator between applications and databases. |
| 6- Very reliable. |

Database Middleware Disadvantages:

| |
|---|
| 1. Does not allow the two-way communications between servers and clients. |
| 2- Cannot transfer calls or objects. |
| 3- Synchronous type of communications can pose problems when multiple demands from multiple users produce huge traffic and congestion. |

## 2.6 AGENT BASED MIDDLEWARE

Agent based middleware are suggested as a solution to solve problems and challenges that arise when creating applications that operate in dynamic, heterogeneous environments, to manage the resources of distributed systems due to the increased flexibility in adapting to the dynamically changing requirements of such systems ( Watson,1996)
. The agent environment provides a set of services shielding agent developers from the low-level details of the underlying platform as depicted in figure 2.

Table 7: Agent Based Middleware Characteristics:

| Property | Value |
| --- | --- |
| Request Reference | Messages |
| Connection Point | Client/Server Cooperative Agent |
| Connection Mode | Negotiation / Synchronous |
| Scalability | High |
| Client state | Unblocked |
| Heterogeneity | High |

Agent Based Middleware Advantages:
1- Modularity, different modules made up by groups of agents may be changed easily.
2- Autonomy and pro-activity properties of agents are well suited for reasoning.
3- Scalability and Adaptability.
4- MAS allows for the interconnection and interoperation of multiple existing legacy systems
( Lin,2004; Fabio,2007)

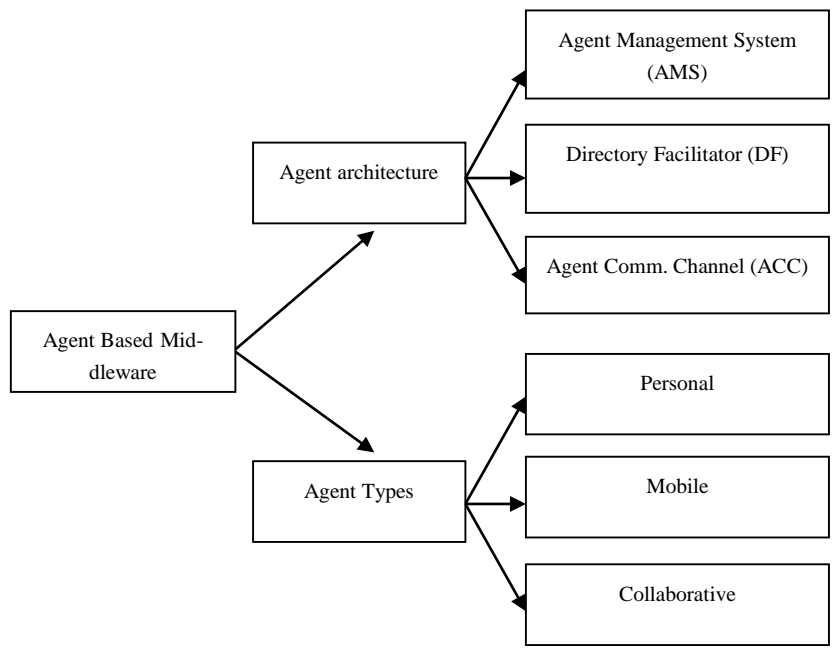

Fig. 2: Agent Based System Classification

*2.7 WEB BASED MIDDLEWARE*

A new approach to system architecture Service- Oriented Architecture (SOA) that was introduced by web services. The SOA can be defined as a software architecture that defines all functions as independent services with well-defined executable interfaces, which can be called in defined sequences to form business processes( Barry,2007). Existing CORBA and DCOM solutions may conform to the above definition, but this is Web services technology that can really allow for wide application integration ( Microsoft Press, 2003). Web services have a collection of underlying technologies as depicted in figure 3:

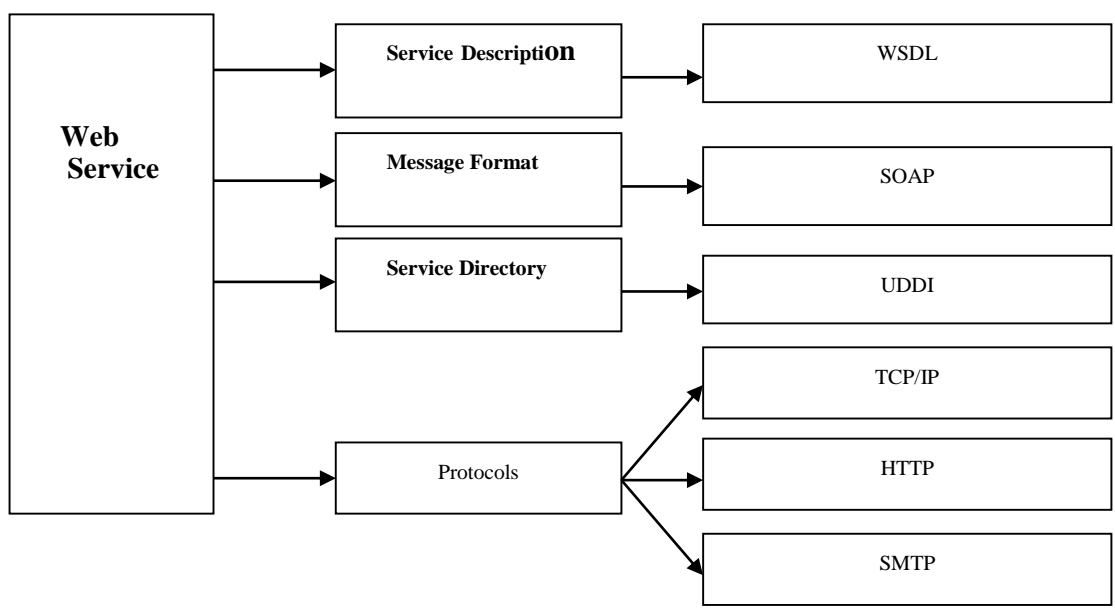

Fig. 3: web services stack

Now after presenting the different types of middleware let's make some comparison of selected features of middleware as appears in table 7:

*3. BUILDING MIDDLEWARE ONTOLOGY*

In recent years' ontologies have become a topic of interest in computer science. Ontology gathers information about certain fields of interest. Ontology describes the concepts in the domain and also the relationships that hold between those concepts. In the next section I will present a middleware ontology structure that can be used for many purposes, including enterprise integration, database design, information retrieval, and information interchanges.

*3.1 - Components of Middleware OWL Ontologies*
An ontology file consists of Individuals, Properties, and Classes ( W3C Recommendation,2019; Matthew,2004; Allemang,2011)

Classes

Classes are interpreted as sets that contain individuals. Classes may be organized into a super-class-subclass hierarchy, which is also known as taxonomy. Subclasses specialize ('are subsumed by') their super-classes. For example consider the classes Middleware and MOM – MOM might be a subclass of Middleware (so Middleware is the super-class of MOM) as depicted in the following figure (figure 4).

Example of classes in our middleware domain is:
{Middleware, Database_Middleware, Call-Type, Connection_Mode

Table 8: comparison between some middleware services

|  | CORBA | DCOM | RMI | EJB | RPC | MOM | WS |
|---|---|---|---|---|---|---|---|
| OS independent | √ |  | √ | √ | √ | √ | √ |
| Languages independent | √ | √ |  |  | √ | √ | √ |
| Data Marshaling | √ | √ | √ | √ | √ |  | √ |
| Synchronous Connection | √ | √ | √ | √ | √ | √ | √ |
| Asynchronous Connection | √ |  |  |  |  | √ | √ |
| Perform Processing | √ | √ | √ | √ | √ |  | √ |
| Make Storage |  |  |  |  |  | √ |  |
| Programmable (Explicit Specs.) | √ | √ | √ | √ | √ |  | √ |

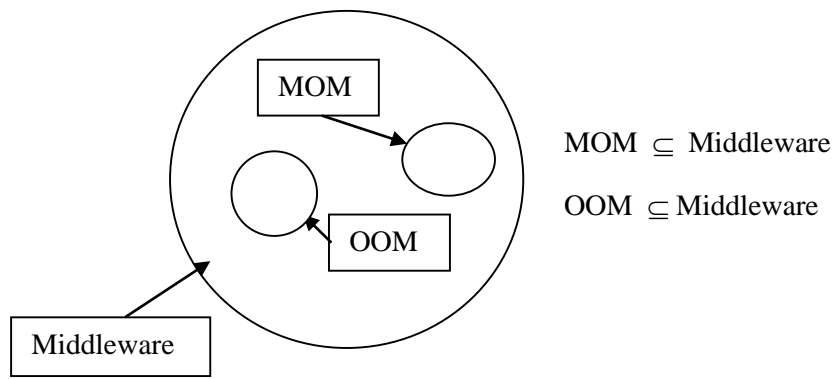

Fig. 4: Representation of classes

Individuals
Individuals, represent objects in the domain that we are interested in. Individuals are also known as instances. Individuals can be referred to as being 'instances of classes'. Figure 5 shows a representation of some individuals in our middleware domain
Example of individuals in our middleware domain is:
{CORBA, DCOM, Web Service, RMI, Synchronous …}

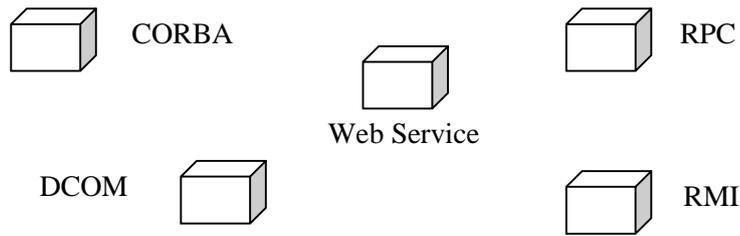

Fig. 5: Representation of Individuals

Properties

There are two main types of properties, Object properties and Datatype properties. Object properties link an individual to an individual. Datatype properties link an individual to an XML Schema Datatype value (xml: integer) or an RDF literal (string). A third type of property known as Annotation properties is also exist, which can be used to add information (metadata) to classes, individuals and object/data type properties. Figure 6 shows a representation of some properties linking some individuals together

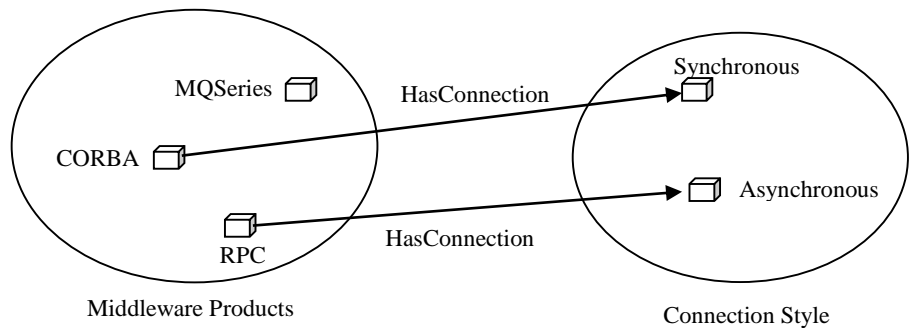

Fig. 6: Representation of Properties

Example of properties in our middleware domain is:
{HasComponent, HasConnection, HasCall …}
Finally, to build our middleware ontology we classify middleware according to different factors as depicted in table 8.

Table 9: Classification of Middleware

| Class | Direct Sub-Classes | Description |
| --- | --- | --- |
| Middleware | Middleware Type (Category) | Contains all classes for different middleware services |
| | Functions | Middleware functionality |
| | Protocols | Different protocols used by middleware services |
| | Call_Type | Methods for calling middleware services |
| | Communication Mode | Sync or Async. |

Our proposed middleware ontology structure as depicted in the following figure 7

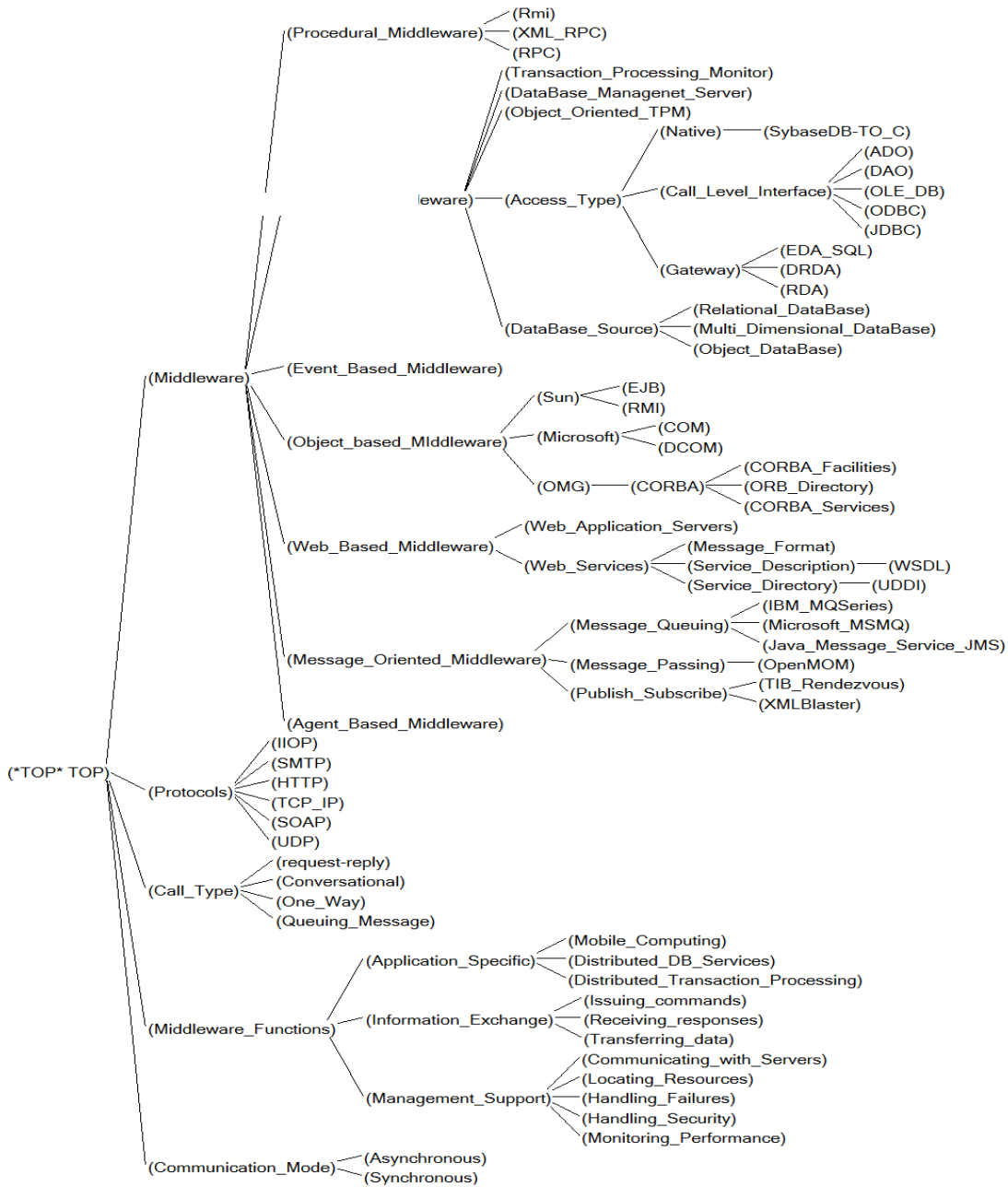

Fig. 7: Middleware Ontology Graph

4 CONCLUSION

In this paper, we have presented a survey about middleware services and their characteristics, as a first step toward building an ontology to list classes and properties of middleware even that some classes in our ontology is not actually a tangible unit that can be parsed as a standalone service, instead it is a middleware function that we could include it in our program to provide a service, e.g. database middleware services, agent middleware.